\documentclass[aps,prl,twocolumn,floatfix,amsmath,amssymb,groupedaddress]{revtex4}
\usepackage[T1]{fontenc}
\usepackage[latin9]{inputenc}
\usepackage{graphicx}
\usepackage{verbatim}
\usepackage{psfrag}
\usepackage[justification=raggedright]{caption}
\makeatletter
\@ifundefined{showcaptionsetup}{}{%
 \PassOptionsToPackage{caption=false}{subfig}}
\usepackage{subfig}
\AtBeginDocument{%
\DeclareRobustCommand*\subref{\@ifstar\sf@@subref\sf@subref}}
\makeatother

\begin{document}

\title{Is dislocation flow turbulent in deformed crystals?}

\author{Woosong Choi}
\email{wc274@cornell.edu}
\author{Yong S. Chen, Stefanos Papanikolaou, James P. Sethna}
\email{sethna@lassp.cornell.edu}
\affiliation{Laboratory of Atomic and Solid State Physics (LASSP),
Clark Hall, Cornell University, Ithaca, New York 14853-2501, USA}

\begin{abstract}
Intriguing analogies were found between a model of plastic deformation in crystals and turbulence in fluids. A study of this model provides remarkable explanations of known experiments and predicts fractal dislocation pattern formation. Further, the challenges encountered resemble those in turbulence, which is exemplified in a comparison with the Rayleigh-Taylor instability.
\end{abstract}
\maketitle

\begin{figure}
	\hfill
	\subfloat[Continuum Dislocation Dynamics]{\label{fig:Dislocations}\includegraphics[width=0.6\columnwidth]{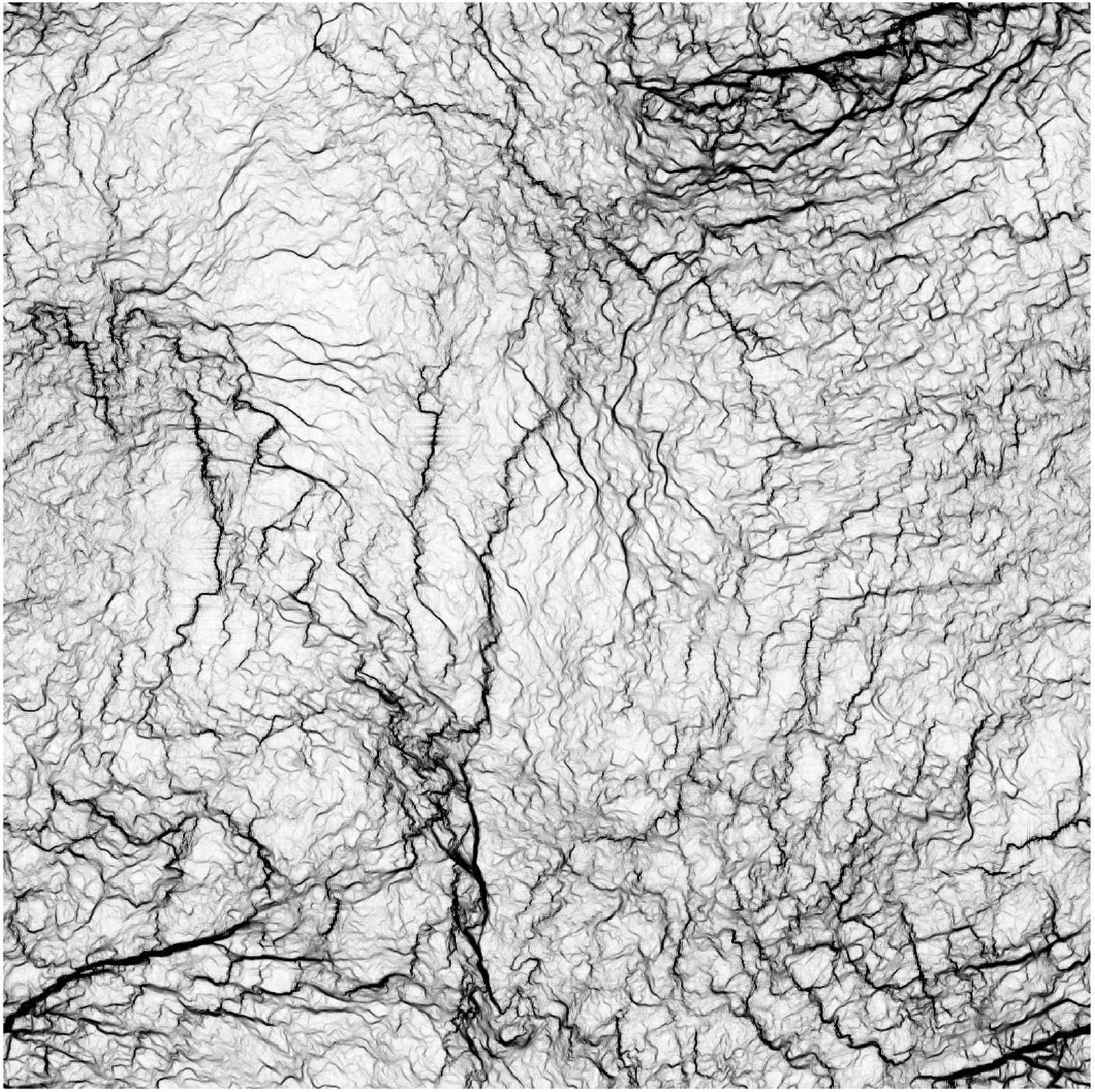}}
	\hfill
	\subfloat[Turbulence]{\label{fig:RTTurbulence}\includegraphics[width=0.3\columnwidth]{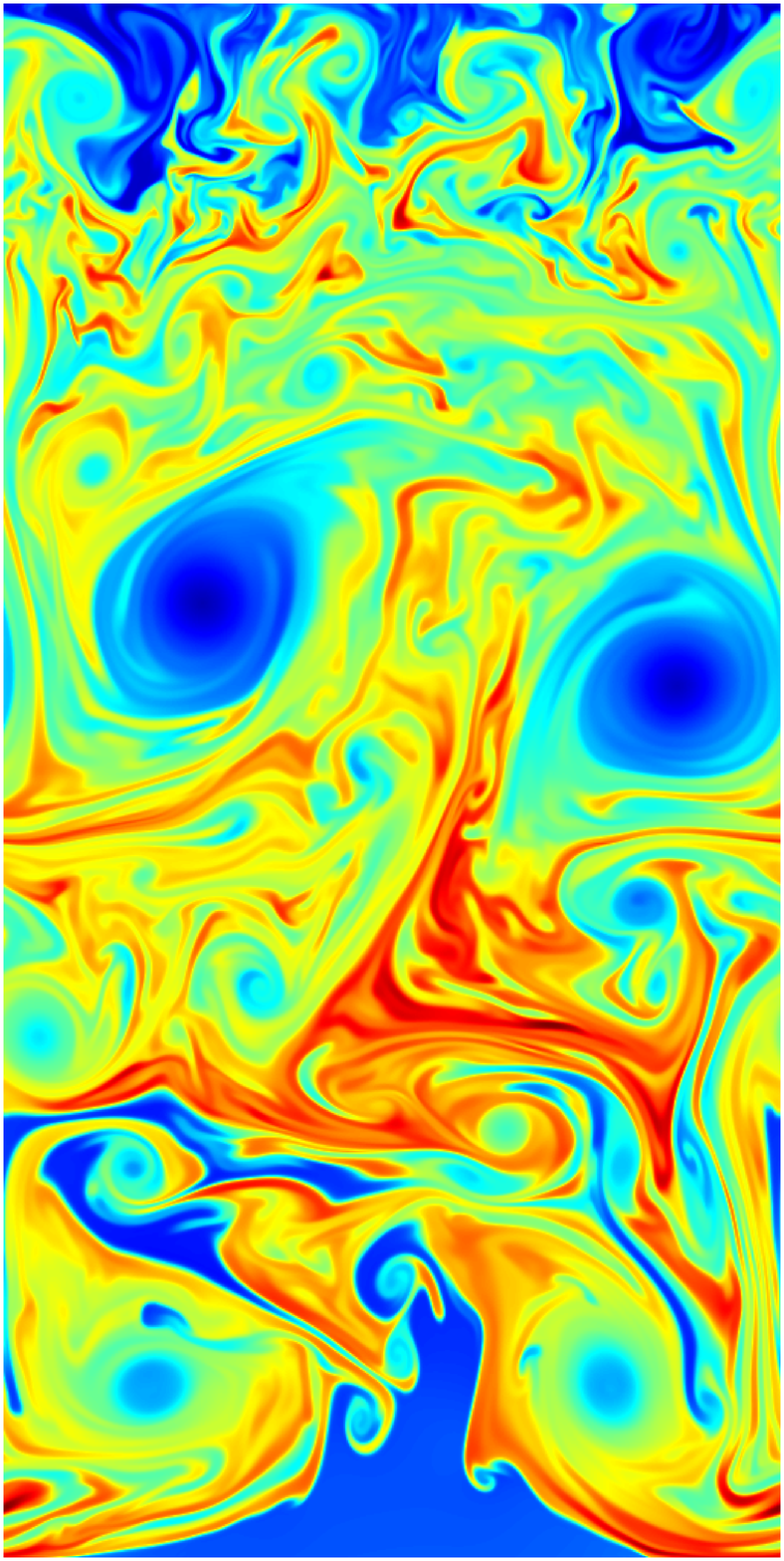}}
	\hfill
	\caption{ {\bf Comparison of our continuum dislocation dynamics~(CDD) with turbulence.} \subref{fig:Dislocations} Dislocation density profile as it evolves from a smooth random initial condition. The structures form fractal {\it cell wall} patterns. Dark regions represent high dislocation density. \subref{fig:RTTurbulence} Rayleigh-Taylor instability at a late time. The fluid~(air) with two layers of different densities mix under the effect of gravity. The emerging flows exhibit complex swirling {\it turbulent} patterns. The color represents density~(red for high, blue for low).}
\label{fig:IntroPictures}
\end{figure}

From horseshoes and knives to bridges and aircrafts, mankind has spent five
millennia studying how the structural properties of metals depend not
only on their constituents, but also how the atoms are arranged and rearranged
as metals are cast, hammered, rolled, and bent into place. A key
part of the physics of this plastic distortion is played by the motion of
intrinsic line defects called {\it dislocations}, and how they move and
rearrange to allow the crystal to change shape. 

Here, we describe the intriguing analogies we found between our model of 
plastic deformation in crystals and turbulence in fluids. Studying this model 
led us to remarkable explanations of existing experiments and let us predict 
fractal dislocation pattern formation. The challenges we encountered resemble 
those in turbulence, which we describe here with a comparison to the 
Rayleigh-Taylor instability.

For brevity, we offer a minimal problem description that ignores many important 
features of plastic deformation of crystals, including yield stress, work 
hardening, dislocation entanglement, and dependence on material 
properties~\cite{Mura1987}. We focus on the complex {\it cellular structures} 
that develop in deformed crystals, which appear to be fractal in some 
experiments~\cite{HahnerPRL1998}. These fractal structures are reproduced 
by our continuum dislocation dynamics (CDD)~\cite{ChenPRL2010} theory (see 
Figure \ref{fig:Dislocations}).

Not only do the resulting patterns match the experimental ones, but the theory 
also has rich dynamics, akin to turbulence. This raises a question: Is the 
dislocation flow turbulent? Here, we focus on exploring this question by 
building analogies to an explicit turbulence example: the Rayleigh-Taylor 
instability. As we describe, our theory displays similar conceptual and 
computational challenges as does this example, which reassures us that we're 
on firm ground.

This CDD model~\cite{LimkumnerdPRL2006,AcharyaJMPS2001,ChenPRL2010} 
provides a deterministic explanation for the emergence of fractal wall 
patterns~\cite{LimkumnerdPRL2006,ChenPRL2010} in mesoscale plasticity.
The crystal's state is described by the
deformation-mediating dislocation density $\varrho_{ij}$ -- where $i$ denotes 
the direction of the dislocation lines and $j$ their Burgers 
vectors~\cite{Mura1987} --
and our dynamical evolution moves this density with a local velocity $V_\ell$,
yielding a partial differential equation~(PDE):
\begin{equation}
  \partial_t \varrho_{ij} - \varepsilon_{imn} \partial_m (\varepsilon_{n\ell k} V_{\ell} \varrho_{kj}) = \nu \partial^4 \varrho_{ij}
  \label{eq:CDD}
\end{equation}
Here $V_\ell$ is
proportional to the net force on it (overdamped motion), coming from the other
dislocations and the external stress. That is, 
\begin{equation*}
V_\ell = \frac{D}{|\varrho|} \sigma_{mn} \epsilon_{\ell mk} \varrho_{kn} 
  \label{eq:V_ell}
\end{equation*}
where $\sigma$ is the local stress tensor, the sum of an external stress
$\sigma^{\rm{ext}}_{ij}$ and the long-range interactions between dislocations.
$\sigma^{\rm{int}}_{ij} = \int K_{ijmn}(r-r') \varrho_{mn}(r') dr'$, with 
$K_{ijmn}$ the function representing the stress at $r$ 
generated by $\varrho$ at $r'$~\cite{Mura1987}. 
The term proportional to $\nu$ is the regularizing quartic diffusion term 
for the dislocation density~(an artificial viscosity), which we'll focus 
on here. (In fact, the equation we simulate here is further
complicated to constrain the motion of the dislocations to the glide plane
while minimizing the elastic energy~\cite{LimkumnerdPRL2006,ChenPRL2010}.)
The details of our equations aren't crucial:
dislocations move around with velocity $\vec{V}$,
pushed by external loads and internal stresses to lower their energies. 
Our equation is
{\it nonlinear}, and it's exactly this non-linearity that makes our theory 
different from more traditional theories of continuum plasticity.

Turbulence is an emergent chaotic flow, typically described by the evolution 
of the Navier-Stokes equations at high {\it Reynolds numbers}:
\begin{equation}
  \begin{array}{r c l}
	\rho\left(\partial_t \vec{v} + \vec{v}\cdot \nabla \vec{v}\right) &=& \mu \nabla^2 \vec{v} + \vec{f}\\
	\partial_t \rho + \nabla \cdot (\rho \vec{v}) &=& 0\\
      \end{array}
  \label{eq:NavierStokes}
\end{equation}
where $\rho$ is the local density of the fluid with velocity $\vec{v}$ under
the application of local external force density $\vec{f}$. The term 
proportional to $\mu$ is the fluid viscosity, and $\mu$ is inversely 
proportional to the Reynolds number.

Despite this Navier-Stokes equation's enormous success  
in describing various experiments, there are many mathematical and
numerical open questions associated with its behavior as $\mu\rightarrow 0$. 
In this regime, complex scale-invariant patterns of eddies and swirls develop
in a way that isn't fully understood: turbulence remains one of the classic 
unsolved problems of science. 

How is $\nu$ related to $\mu$? Eq.\,(\ref{eq:NavierStokes}) can be written 
differently by dividing the whole equation by $\rho$, in this equation 
$\mu/\rho$~(in the incompressible case) is called the 
{\it kinematic viscosity}~(usually denoted as $\nu$). Our artificial viscosity
$\nu$ in Eq.\,(\ref{eq:CDD}) is analogous to this kinematic viscosity. 
For turbulence, $\mu$ in Eq.\,(\ref{eq:NavierStokes}) is given by nature. 
In contrast, our $\nu$ in Eq.\,(\ref{eq:CDD}) is added for numerical stability; 
it smears singular walls to give regularized solutions. This is physically 
justified because the atomic lattice always provides a cutoff scale. 
How do we know that this artificial term gives the `correct' answer~(given 
that there can be many different solutions to the same PDE)?
Numerical methods for shock-admitting PDEs are validated by showing that the 
vanishing grid spacing limit $h\to0$ gives the same solution as the $\nu\to0$ 
limit~(that is, the {\it viscosity solution}).
We will argue that both our model of plasticity and the Navier-Stokes equations 
do not have convergent solutions as $\mu$ or $\nu\to0$. 
We're reassured from the turbulence analogy and are satisfied with extracting 
physically sensible results from our plasticity theory
-- viewing it not as {\it the} theory of plasticity, but as an {\it acceptable} 
theory.

To solve Eq.\,(\ref{eq:CDD}), we implemented a second order central upwind 
method~\cite{KurganovSIAMJSC2002} especially developed and tested for 
conservation laws, such as Eqs.\,(\ref{eq:CDD})~and~(\ref{eq:NavierStokes}). 
The method uses a generalized approximate Riemann solver which doesn't demand 
the full knowledge of characteristics~\cite{KurganovSIAMJSC2002}. 
For the simulations of Navier-Stokes dynamics (Eq.\,(\ref{eq:NavierStokes})) we 
use PLUTO~\cite{MignoneApJS2007}, a software package built to run hydrodynamics 
and magnetohydrodynamics simulations, using the Roe approximate Riemann solver.

\begin{figure}

\subfloat[Non-convergence in dislocation dynamics~(see Eq.\,(\ref{eq:CDD}))]{\label{fig:NonconvergenceCDD}\includegraphics[width=0.9\columnwidth]{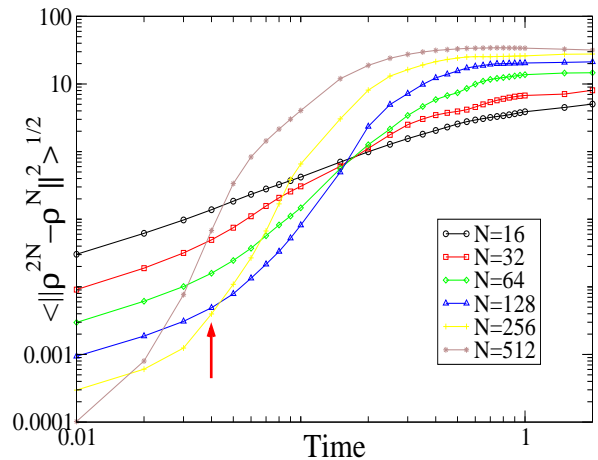}}

\subfloat[Non-convergence in turbulence dynamics~(Rayleigh-Taylor instability)]{\label{fig:NonconvergenceRT}\includegraphics[width=0.9\columnwidth]{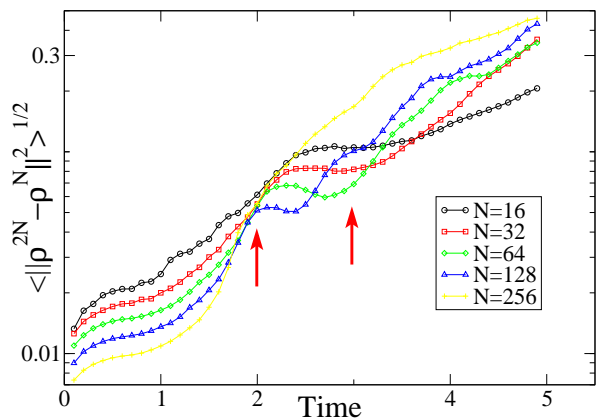}}

\caption{ {\bf Non-convergence exhibited in both plasticity and turbulence.} As time progresses, the curves, which are initially monotonically decreasing, flip order and become non-convergent~(where the lines cross each other). Red arrows show where the convergence is lost.}
\end{figure}

Accurately capturing singular flows is a challenge in computational fluid 
dynamics. A classic example of such singularities is the sonic
boom that happens when an object passes through a compressible fluid
(described by a version of Navier-Stokes) faster than its speed of sound.
The sonic boom is a sharp jump in density and pressure, which causes
the continuum equations to become ill-defined. Our PDEs,
depending on gradients of $\varrho$, become ill-defined when $\varrho$
develops an infinite gradient at a dislocation density jump. 

The numerical methods we use are designed to appropriately solve the so-called 
Riemann problem: the evolution of a simple initial condition with a single step 
in the conserved physical quantities.
For hyperbolic conservation laws, exact solutions of the Riemann problem can be 
obtained by decomposing the step into characteristic waves. However, in most non
linear problems, finding exact Riemann solutions involves iterative processes th
at are either slow or (practically) impossible, and thus approximate solutions 
are employed instead. Both methods we use are approximate in different ways, 
but are qualitatively similar. 

\begin{figure*}
  \hfill{}\subfloat[$h=1/128$]{\includegraphics[width=0.3\textwidth]{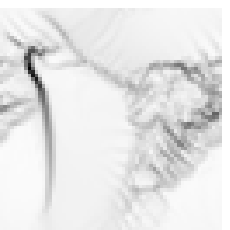}}
  \hfill{}\subfloat[$h=1/256$]{\includegraphics[width=0.3\textwidth]{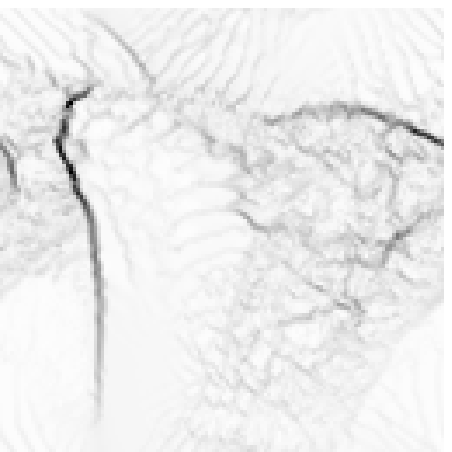}}
  \hfill{}\subfloat[$h=1/512$]{\includegraphics[width=0.3\textwidth]{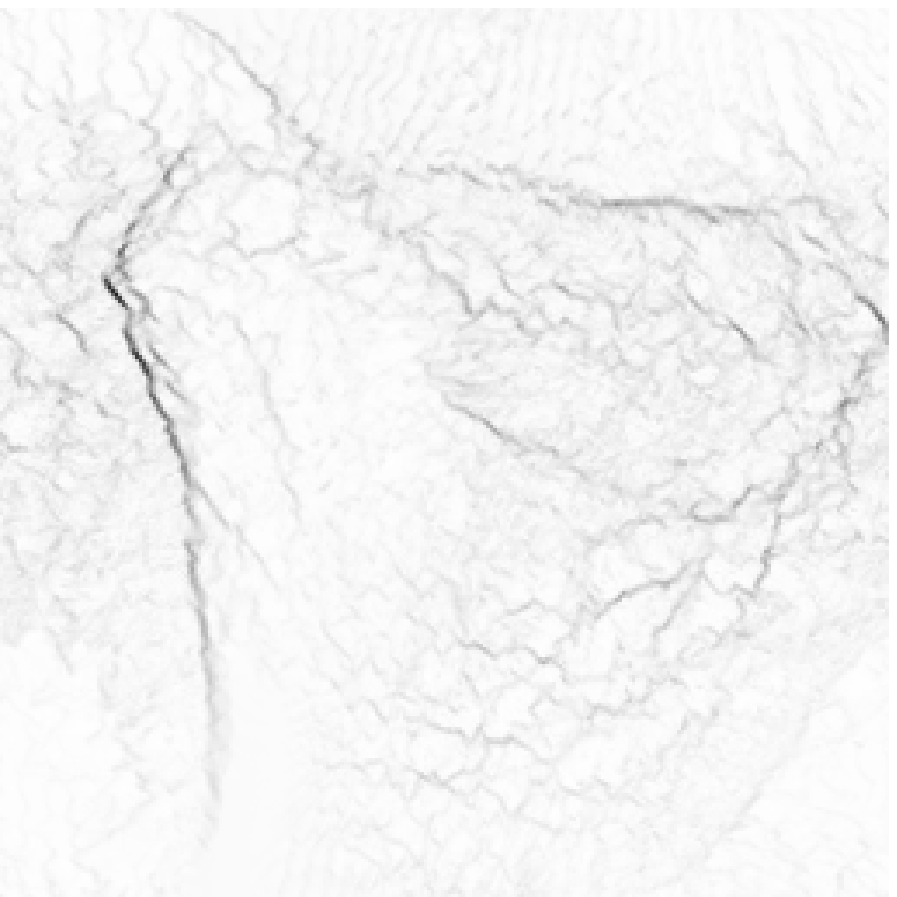}}
  \hfill{}

  \caption{ {\bf Continuum dislocation dynamics.} Simulation results at $t=1.0$ of our CDD Eq.\,(\ref{eq:CDD}) at different grid sizes~($h$), starting from a smooth initial condition. We use periodic boundaries in both horizontal and vertical directions, and all physical quantities are constant along the perpendicular direction.}
\label{fig:CDDSizes}
\end{figure*}

\begin{figure}
  \hfill{}\subfloat[$h=1/128$]{\label{subfig:RT_128}\includegraphics[width=0.3\columnwidth]{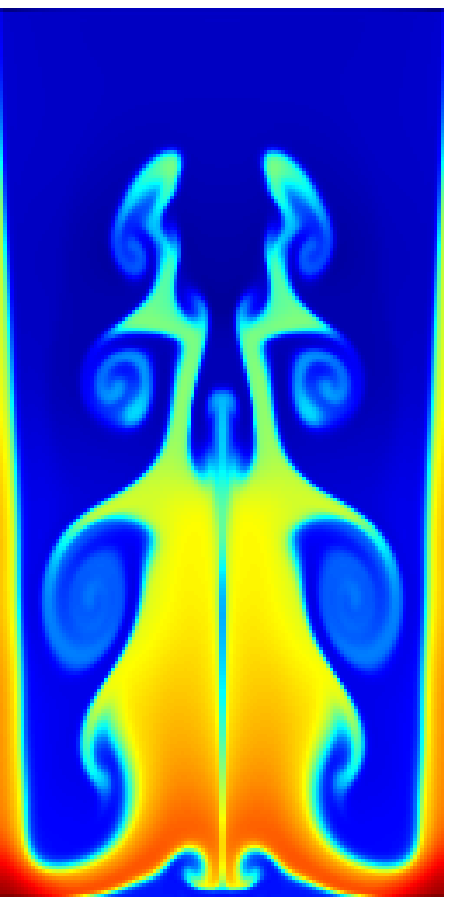}}
  \hfill{}\subfloat[$h=1/256$]{\label{subfig:RT_256}\includegraphics[width=0.3\columnwidth]{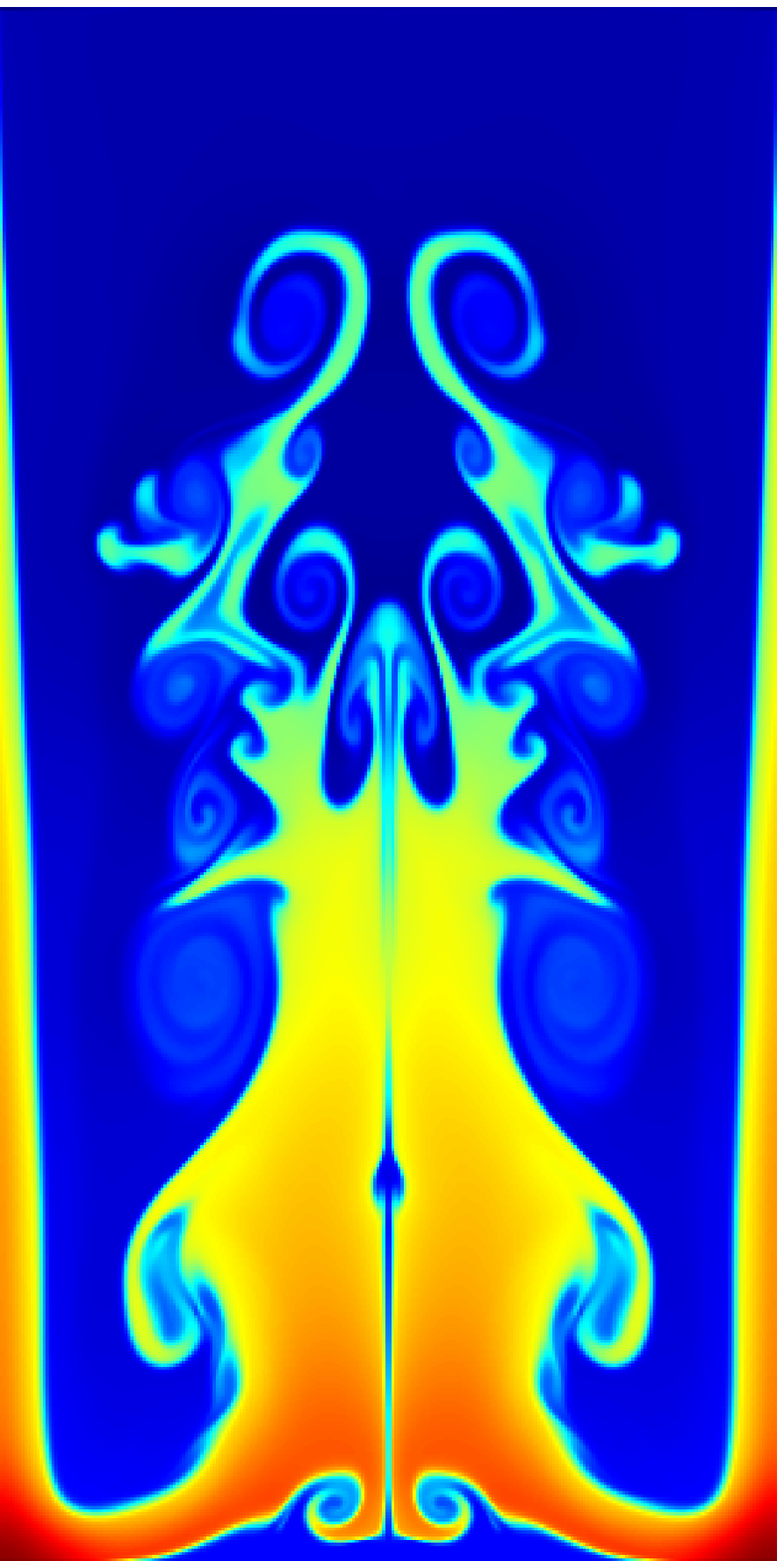}}
  \hfill{}\subfloat[$h=1/512$]{\label{subfig:RT_512}\includegraphics[width=0.3\columnwidth]{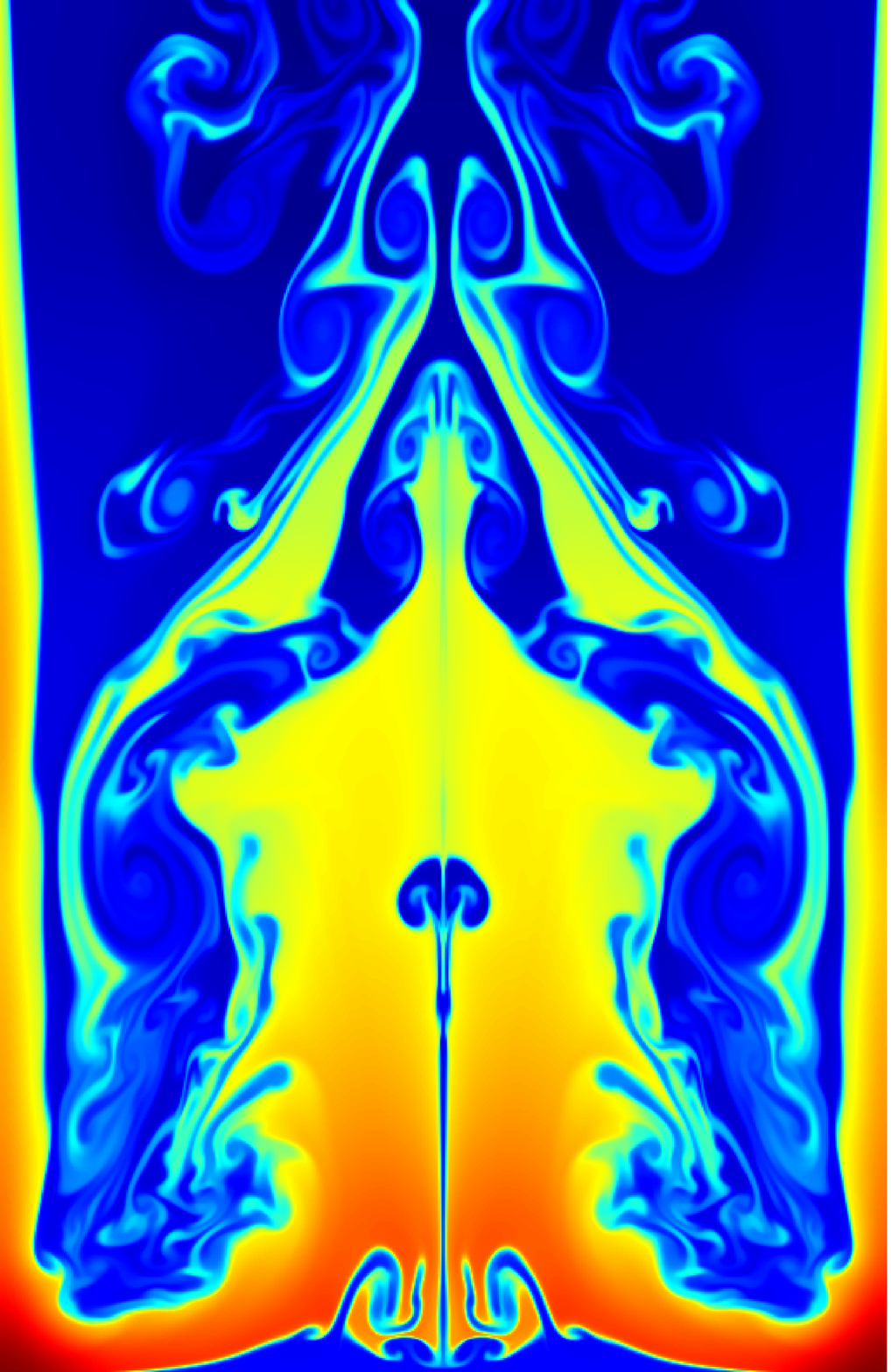}}
\hfill{}

\caption{ {\bf The Rayleigh-Taylor instability of the Navier-Stokes equation.} The Rayleigh-Taylor instability is a fluid mixing phenomenon that occurs when an interface between two different fluid densities is pulled by gravity. These simulation results here are the Rayleigh-Taylor instability at $t=4.0$, for $\mu \to 0$, at different grid spacings~($h$) with periodic boundaries in the horizontal direction and fixed boundaries along the vertical. The initial condition has density interface with a single mode perturbation in the vertical velocity. The system size is $(L_x,L_y) = (1.0,2.0)$.}
\label{fig:RTNavierStokes}
\end{figure}

These sophisticated methods are designed to handle the kind of density jumps
seen in sonic booms. In our dislocation dynamics, though, we have a more severe
singularity that forms -- a sharp wall of dislocations that 
becomes a $\delta$-function singularity in the dislocation 
density~(as $\nu\to0$). These $\delta$-shocks are naturally present in 
crystals -- they describe, for example, the grain boundaries found in 
polycrystalline metals, which (in the continuum limit) form sharp walls of 
dislocations separating dislocation-free crystallites. 
Unfortunately, the mathematical and computational understanding of
PDEs forming $\delta$-shocks is relatively primitive;
there are only a few analytic and numerical studies in one dimension~(for 
an example, see~\cite{TanJDE1994}).
Currently, to our knowledge, there's no numerical method especially designed 
for $\delta$-shock solutions.
Moreover, in a strict mathematical sense, several properties of 
Eq.\,(\ref{eq:CDD}) -- nonlocality and mixed hyperbolic and parabolic features 
-- haven't been proven to permit a successful application of shock-resolving 
numerical methods.

In the simulations presented here, we won't add an explicit viscosity~(so 
$\mu=\nu=0$); instead, we have an effective numerical 
dissipation~\cite{KurganovSIAMJSC2002} that depends on the grid spacing $h$ 
as $h^n$, where $n$ depends on the numerical method 
used.~(Eq.\,(\ref{eq:NavierStokes}) with $\mu=0$ is the compressible Euler 
equation. Although we present our simulations as small numerical viscosity 
limits of Navier-Stokes, they could be viewed as particular approximate 
solutions of these Euler equations.)

Figure \ref{fig:IntroPictures} shows typical emerging structures in simulations 
of both the CDD Eq.\,(\ref{eq:CDD}) and the Navier-Stokes dynamics 
Eq.\,(\ref{eq:NavierStokes}): both are complex, displaying structures at many 
different length scales; sharp, irregular walls in the CDD and vortices in 
Navier-Stokes.
Although it might not be surprising to professionals in fluid mechanics that 
nonlinear PDEs have complex, self-similar solutions, it's quite startling to 
those studying plasticity that their theories can contain such complexity~(even 
though this complexity has been observed in experiments~\cite{HahnerPRL1998}): 
traditional plasticity simulations do not lead to such structures.

These rich and exotic solutions demand scrutiny. How do we confirm the validity 
of our solutions?
For continuum PDEs solved on a grid, an important problem that needs to be 
addressed is the effect of the imposed grid. 
Traditionally, it's expected that as the grid becomes finer, the solution
is likely to be closer to the real continuum solution. 
For differential equations that generate singularities, one cannot expect 
simple convergence at the singular point! How do we define convergence when 
singularities are expected?
For ordinary density-jump shocks like sonic booms, mathematicians have
defined the concept of a {\it weak solution}: it's a solution to the 
integrated version of the original equation, bypassing singular derivatives. 

For many problems, researchers have shown that adding an artificial viscosity 
and taking the limit to zero yields a weak solution to the problem. 
For some problems, the weak solution is unique, while for others there 
can be several:
different numerical methods or types of regularizing viscosities
can yield different dynamics of the singularities. This makes physical sense:
if a singular defect (a dislocation or a grain boundary) is defined on an
atomic scale, shouldn't the details of how the atoms move (ignored in the 
continuum theory) be important for the defect's motion? In the 
particular case of sonic booms, the microscopic physics picks out the
{\it viscosity solution} (given by an appropriate $\mu\to0$ limit), leading
mathematicians to largely ignore the question of how micro-scale physics 
determines the singularities' motion.

However, our problems here are more severe than picking
out a particular weak solution. Neither our dislocation dynamics nor the
Navier-Stokes equation~(with very high Reynolds number) converge 
in the continuum limit even for gross features,
whether we take the grid size to zero in the upwind schemes or we take
$\nu\to0$~(or $\mu\to0$) as a mathematical limit. 

Figure~\ref{fig:NonconvergenceCDD} shows a quantitative measure of the
our simulation's convergence as a function of time, as the grid spacing 
$h=1/N$ becomes smaller.
We measure convergence using the $L_2$ norm 
\begin{equation*}
||\widehat{\varrho^{2N}}-\varrho^{N}||_2 \equiv \left(\int \Vert \widehat{\varrho^{2N}}-\varrho^{N} \Vert^2 \,dx\right)^{1/2}
  \label{eq:L2norm}
\end{equation*}
where $\widehat{\varrho^{2N}}$ has been suitably smeared to the resolution of 
$\varrho^N$.
(Normally we'd check the difference between the current solution and
the true answer, but here we don't know the true answer.) Here, We study the 
relaxation of a smooth but randomly chosen initial
condition -- that is, a perfect single crystal beaten with mesoscale hammers 
with round heads -- as a function of time.  We see that for short times these 
distances converge rapidly to zero, implying convergence of our solution in 
the $L_2$ norm.  However, at around $t=0.02$ to $0.2$, the solutions 
begin to become increasingly different as the grid spacing $h\to0$.

This worried us at the beginning because it suggests that the numerical
results might be dependent somehow on the artificial finite-difference grid
we use to discretize the problem,
and therefore might not reflect the correct continuum physical solution.
We checked this by adding the aforementioned artificial viscosity $\nu$ in
Eq.\,(\ref{eq:CDD}). 
We found that it converges nicely when $\nu$ is fixed
as the grid spacing goes to zero. However, this converged solution is
not unique: it keeps changing as $\nu\to0$.
So, it's our fundamental equation
of motion (Eq.\,(\ref{eq:CDD})) and not our numerical method that fails to 
have a continuum solution. This would seem even more worrisome: 
How do we understand
a continuum theory whose predictions seem to depend on the smallest studied 
length scale (the atomic size)?

It's here that the analogy to turbulence has been crucial for understanding
the physics. 
It's certainly not obvious that the limit of strong turbulence
$\mu\to0$ in Navier-Stokes (Eq.\,(\ref{eq:NavierStokes})) should converge to
a limiting flow. 
Actually, our short experience suggests that there is no viscosity solution 
for Eq.\,(\ref{eq:NavierStokes}).
Turbulence has a hierarchy of eddies and swirls on all
length scales, and as the viscosity decreases (for fixed initial conditions
and loading) not only do the small-scale eddies get smaller, but also 
the position of the large-scale eddies at fixed time change as the viscosity or grid size is reduced.

Figure \ref{fig:NonconvergenceRT} depicts the convergence behavior of a
simulation of the Rayleigh-Taylor instability~(as in
Fig.~\ref{fig:NonconvergenceCDD}). 
The instability triggers turbulent flow, and like our dislocation simulations, 
convergence is lost after $t\sim 2.0$ as the grid spacing $h$ gets smaller.
Our choice of the Rayleigh-Taylor instability for comparison is motivated by 
the presence of robust self-similar features~(such as the ``bubbles'' and 
``spikes'' in Fig.~\ref{fig:RTNavierStokes}~\cite{ShiJCP2003}), and by the 
spatio-temporally non-converging features of the initially well-defined 
interface. 

The interface between two fluid densities is analogous to our dislocation 
cell walls. Even though the Rayleigh-Taylor instability is different from 
homogeneous turbulence in important ways, we also verified that the latter 
shows similar spatio-temporal non-convergence but statistical 
convergence~(simulating the Kelvin-Helmholtz instability for compressible flow 
in 2D). The Rayleigh-Taylor instability provides the best visualization of the 
analogy between the two phenomena, but this non-convergence appears to be more 
general.
Fig.~\ref{fig:RTNavierStokes} shows density profiles at intermediate times 
for the Rayleigh-Taylor instability. The figure shows the formation of 
vortices~(swirling patterns), and the simulations look significantly different 
as the grid spacing decreases.  Again, this is analogous to the corresponding 
simulations of our dislocation dynamics shown in Fig.~\ref{fig:CDDSizes}, 
where larger cells continue to distort and shift as the grid spacing decreases.

\begin{figure}
\includegraphics[width=0.9\columnwidth]{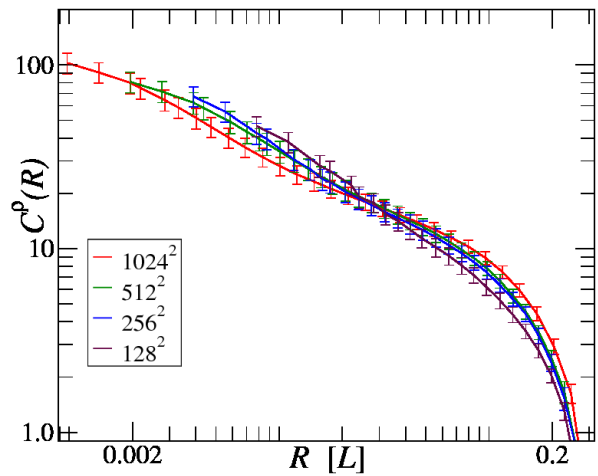}
\caption{ {\bf Statistical properties and convergence} Although the continuum dislocation dynamics~(CDD) simulations with different grid sizes are non-convergent~(Fig.~\ref{fig:NonconvergenceCDD}), the statistical properties are the same. The dislocation density correlation function is plotted here for different simulation sizes at the same time, exhibiting consistent power laws.}
\label{fig:StatisticalConvergence}
\end{figure}

If the simulations aren't convergent, how can we decide if the theory
is physically relevant and can be trusted to interpret experiments?
In turbulence, it has long been known that, as vortices develop, self-similar 
patterns arise in the flow and exhibit power laws in the energy 
spectrum and in the velocities' correlation functions~\cite{Frisch1995}. 
A successful simulation of fully developed turbulence isn't judged by whether 
the flow duplicates an exact solution of Navier-Stokes! Turbulence simulations 
study these power laws, comparing them to analytical predictions and
experimental measurements.

Our primary theoretical focus in our plasticity study~\cite{ChenPRL2010} has 
been to analyze power-law correlation functions for the dislocation density, 
plastic distortion tensor, and local crystalline orientation.
As Fig.~\ref{fig:StatisticalConvergence} shows, like turbulence simulations, these statistical properties seem to converge nicely in the continuum limit.

\begin{figure*}
  \hfill{}\subfloat[$h=1/256$]{\fbox{\label{subfig:256_0_04}\includegraphics[width=0.45\columnwidth]{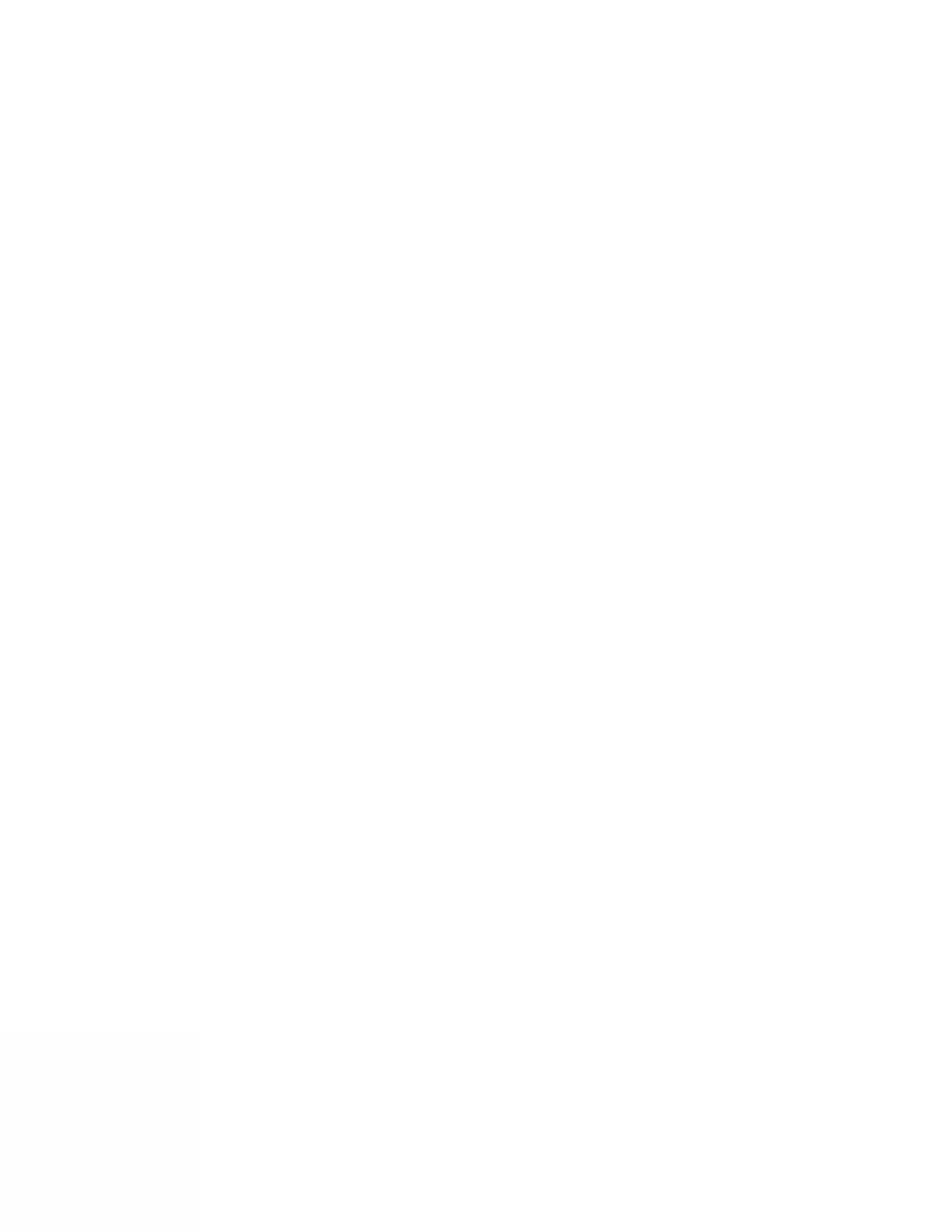}}}
  \hfill{}\subfloat[$h=1/512$]{\fbox{\label{subfig:512_0_04}\includegraphics[width=0.45\columnwidth]{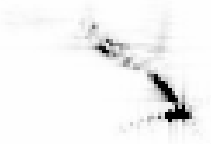}}}
  \hfill{}\vline
  \hfill{}\subfloat[$h=1/256$]{\fbox{\label{subfig:256_1_00}\includegraphics[width=0.45\columnwidth]{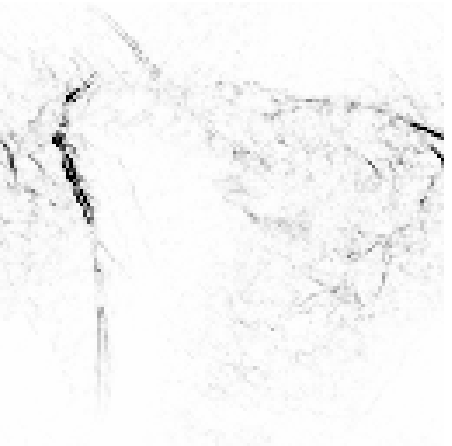}}}
  \hfill{}\subfloat[$h=1/512$]{\fbox{\label{subfig:512_1_00}\includegraphics[width=0.45\columnwidth]{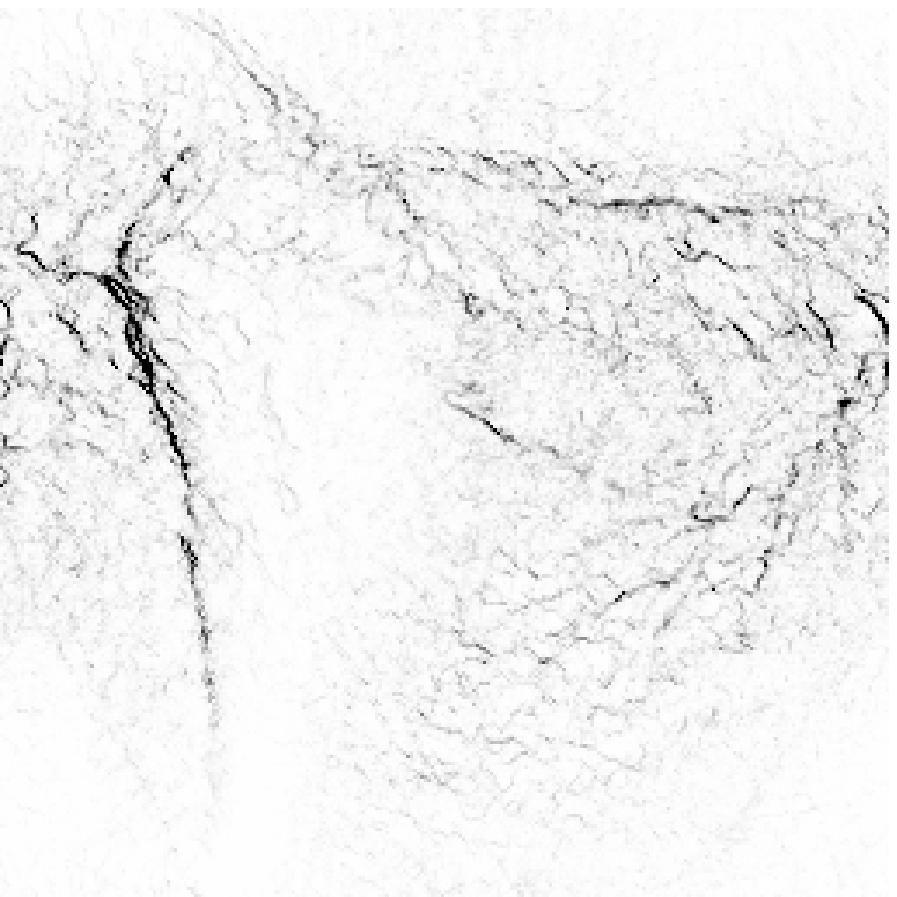}}}

  \vspace{.05in}
  \hfill{}$t=0.04$\hfill{}
  \hfill{}$t=1.00$\hfill{}
  \caption{ {\bf Non-convergence and singularity for CDD} For \subref{subfig:256_0_04} and \subref{subfig:512_0_04}, $t=0.04$; for \subref{subfig:256_1_00} and \subref{subfig:512_1_00}, $t=1.00$. The two-norm difference between $h$ and $h/2$ are plotted. At short time, $t=0.04$, the differences are small: \subref{subfig:256_0_04} is empty and \subref{subfig:512_0_04} nearly so. At later times, $t=1.00$, the two-norm difference becomes significant esepcially where the walls are forming~(see Fig.~\ref{fig:CDDSizes} for wall locations).}
  \label{fig:NonconvergenceSingularityCDD}
\end{figure*}

\begin{figure*}
  \hfill{}\subfloat[$h=1/64$]{\fbox{\label{subfig:64_2.0}\includegraphics[width=0.3\columnwidth]{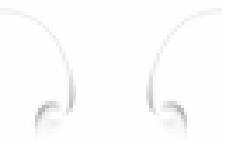}}}
  \hfill{}\subfloat[$h=1/128$]{\fbox{\label{subfig:128_2.0}\includegraphics[width=0.3\columnwidth]{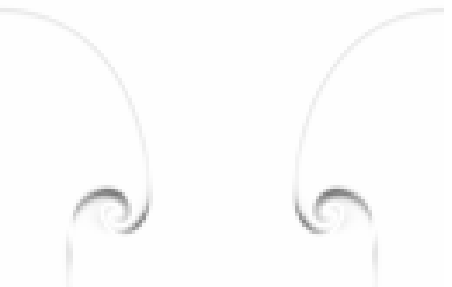}}}
  \hfill{}\subfloat[$h=1/256$]{\fbox{\label{subfig:256_2.0}\includegraphics[width=0.3\columnwidth]{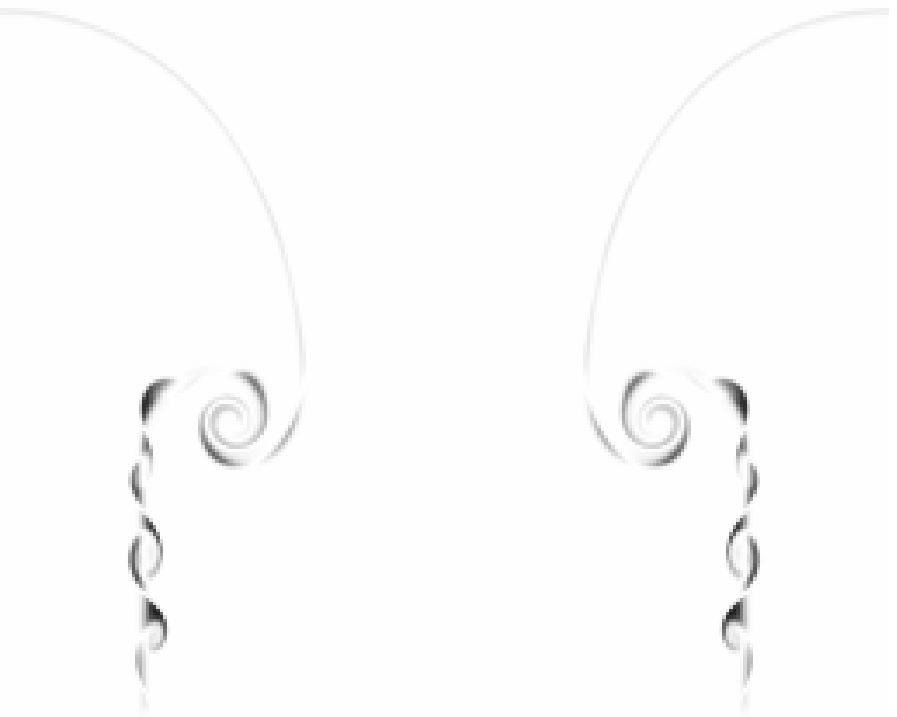}}}
  \hfill{}\vline
  \hfill{}\subfloat[$h=1/64$]{\fbox{\label{subfig:64_3.0}\includegraphics[width=0.3\columnwidth]{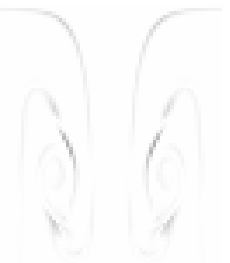}}}
  \hfill{}\subfloat[$h=1/128$]{\fbox{\label{subfig:128_3.0}\includegraphics[width=0.3\columnwidth]{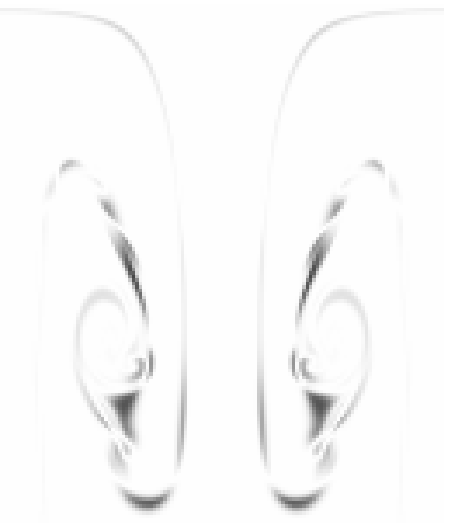}}}
  \hfill{}\subfloat[$h=1/256$]{\fbox{\label{subfig:256_3.0}\includegraphics[width=0.3\columnwidth]{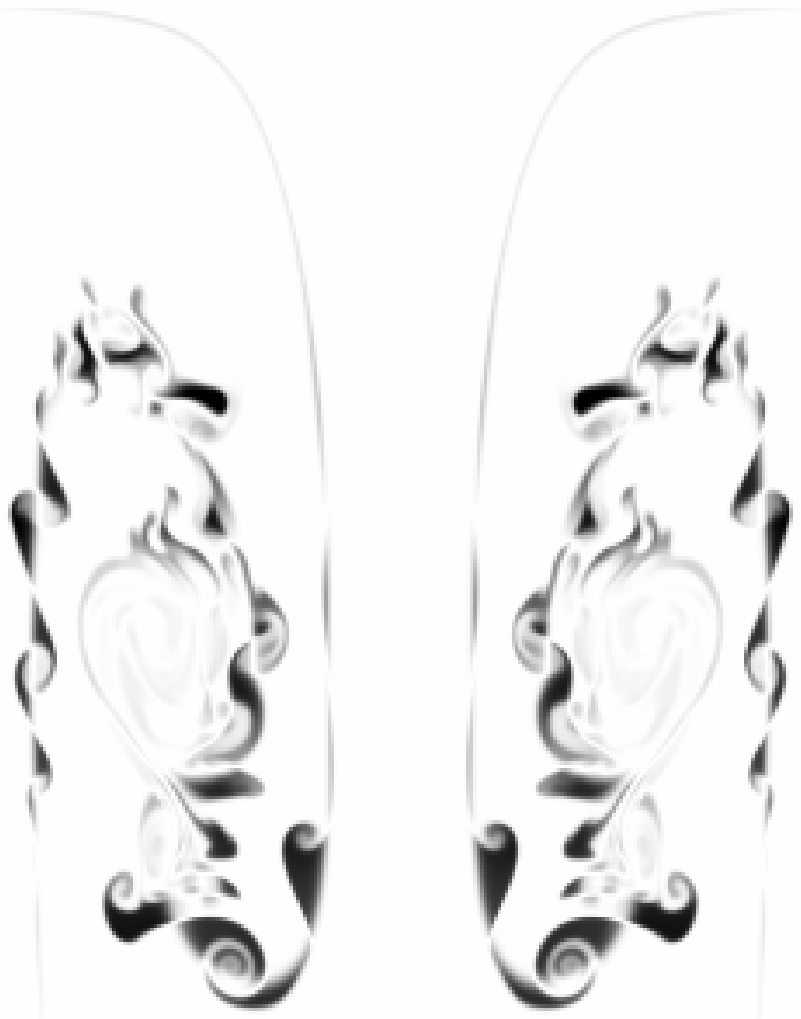}}}

  \vspace{.05in}
  \hfill{}$t=2.0$\hfill{}
  \hfill{}$t=3.0$\hfill{}
  \hfill{}\caption{ {\bf Non-convergence and vortices for Navier-Stokes} For \subref{subfig:64_2.0}, \subref{subfig:128_2.0}, and \subref{subfig:256_2.0} $t=2.0$; for \subref{subfig:64_3.0}, \subref{subfig:128_3.0}, and \subref{subfig:256_3.0} $t=3.0$, corresponding to the two red arrows in Figure~\ref{fig:NonconvergenceRT}. The two-norm difference between $h$ and $h/2$ is plotted. At $t=2.0$, small structures start to become strong, as in \subref{subfig:256_2.0}, and the same is true at $t=3.0$ for $h=1/128$, as in \subref{subfig:128_3.0}.}
  \label{fig:NonconvergenceSingularity}
\end{figure*}

It's worth noting that, in both cases, non-convergence emerges when small scale 
features appear on the wall~(see Fig.~\ref{fig:NonconvergenceSingularityCDD}) 
or the interface~(see Fig.~\ref{fig:NonconvergenceSingularity}):

In the case of our simulations of plastic flow~(Eq.\,\ref{eq:CDD}), starting 
from smooth initial density profiles, finite time singularities develop in 
the form of $\delta$-shocks. The existence of finite-time singularities
was shown in a 1D variant of these equations, which is associated with the Burgers equation~\cite{LimkumnerdJMPS2007}.
Figure~\ref{fig:NonconvergenceSingularityCDD} shows how this effect occurs by 
considering the two-norm differences~(the integrand in space of the $L_2$ norm 
discussed earlier). At $t=0.04$ (see Fig.~\ref{fig:NonconvergenceCDD}) 
when the $N=512$ curve starts to cross all the other curves, 
singular features start to appear around a wall~(Fig.~\ref{subfig:512_0_04}). 
Although the boundaries are non-convergent when
specific locations and times are considered~(Fig.~\ref{fig:NonconvergenceCDD}), the statistical properties and associated 
self-similarity~(Fig.~\ref{fig:StatisticalConvergence}) are convergent.

In the case of Navier-Stokes simulations~(see 
Fig.~\ref{fig:NonconvergenceSingularity}), the existence of finite-time
singularities is a topic of active research: even though local-in-time analytic 
solutions are easily shown to exist, global-in-time analytic solutions can be 
proven to exist only for special cases, such as in the 2D incompressible
genuine Euler equation~\cite{BardosArxiv2010}. 
In 3D, the mechanism of vortex stretching is conjectured to lead to finite-time
singularities~\cite{PumirPRL1992}, even though there are still crucial open questions. Despite its complexity, turbulence
can be concretely studied in special cases. For our example of the 
Rayleigh-Taylor instability, the two-fluid interface gets distorted and 
``bubbles'' form (Fig.~\ref{fig:RTNavierStokes}); over time, the bubbles 
exhibit emergent self-similar characteristics~\cite{ShiJCP2003}, showing 
{\it statistical convergence}. However, there is no spatio-temporal convergence,
because the interface develops complex, turbulent features as the grid becomes 
finer~(see Figs.~\ref{fig:NonconvergenceRT} and 
\ref{fig:NonconvergenceSingularity}).

Sometimes science seems to be fragmented, with independent fields whose 
vocabularies, toolkits, and even philosophies almost completely separate. 
But many valuable insights and advances arise when ideas from
one field are linked to another. Computational science is providing a new
source of these links, by tying together fields that can fruitfully share
numerical methods. 

Our use of well-established numerical methods from the fluids community made 
it both natural and easy to utilize their analytical methods for judging 
the validity of our simulations and interpreting their results.

We thank Alexander Vladimirsky, Randall LeVeque, Chi-Wang Shu, and Eric Siggia 
for helpful and inspiring discussions on numerical methods and turbulence. The 
US Department of Energy/Basic Energy Sciences grant no. DE-FG02-07ER46393 
supported our work, and the US National Center for Supercomputing Applications 
partially provided computational resources on the Lincoln and Abe clusters 
through grant no. MSS090037.


\begin{thebibliography}{13}
\expandafter\ifx\csname natexlab\endcsname\relax\def\natexlab#1{#1}\fi
\expandafter\ifx\csname bibnamefont\endcsname\relax
  \def\bibnamefont#1{#1}\fi
\expandafter\ifx\csname bibfnamefont\endcsname\relax
  \def\bibfnamefont#1{#1}\fi
\expandafter\ifx\csname citenamefont\endcsname\relax
  \def\citenamefont#1{#1}\fi
\expandafter\ifx\csname url\endcsname\relax
  \def\url#1{\texttt{#1}}\fi
\expandafter\ifx\csname urlprefix\endcsname\relax\def\urlprefix{URL }\fi
\providecommand{\bibinfo}[2]{#2}
\providecommand{\eprint}[2][]{\url{#2}}

\bibitem[{\citenamefont{Mura}(1987)}]{Mura1987}
\bibinfo{author}{\bibfnamefont{T.}~\bibnamefont{Mura}},
  \emph{\bibinfo{title}{Micromechanics of defects in solids}}
  (\bibinfo{publisher}{Springer}, \bibinfo{year}{1987}), \bibinfo{edition}{2nd} ed.

\bibitem[{\citenamefont{H\"{a}hner et~al.}(1998)\citenamefont{H\"{a}hner, Bay,
  and Zaiser}}]{HahnerPRL1998}
\bibinfo{author}{\bibfnamefont{P.}~\bibnamefont{H\"{a}hner}},
  \bibinfo{author}{\bibfnamefont{K.}~\bibnamefont{Bay}}, \bibnamefont{and}
  \bibinfo{author}{\bibfnamefont{M.}~\bibnamefont{Zaiser}},
  \bibinfo{journal}{Phys. Rev. Lett.} \textbf{\bibinfo{volume}{81}},
  \bibinfo{pages}{2470} (\bibinfo{year}{1998}).

\bibitem[{\citenamefont{Chen et~al.}(2010)\citenamefont{Chen, Choi,
  Papanikolaou, and Sethna}}]{ChenPRL2010}
\bibinfo{author}{\bibfnamefont{Y.~S.} \bibnamefont{Chen}},
  \bibinfo{author}{\bibfnamefont{W.}~\bibnamefont{Choi}},
  \bibinfo{author}{\bibfnamefont{S.}~\bibnamefont{Papanikolaou}},
  \bibnamefont{and} \bibinfo{author}{\bibfnamefont{J.~P.}
  \bibnamefont{Sethna}}, \bibinfo{journal}{Phys. Rev. Lett.}
  \textbf{\bibinfo{volume}{105}}, \bibinfo{pages}{105501}
  (\bibinfo{year}{2010}).

\bibitem[{\citenamefont{Limkumnerd and Sethna}(2006)}]{LimkumnerdPRL2006}
\bibinfo{author}{\bibfnamefont{S.}~\bibnamefont{Limkumnerd}} \bibnamefont{and}
  \bibinfo{author}{\bibfnamefont{J.~P.} \bibnamefont{Sethna}},
  \bibinfo{journal}{Phys. Rev. Lett.} \textbf{\bibinfo{volume}{96}},
  \bibinfo{pages}{095503} (\bibinfo{year}{2006}).

\bibitem[{\citenamefont{Acharya}(2001)}]{AcharyaJMPS2001}
\bibinfo{author}{\bibfnamefont{A.}~\bibnamefont{Acharya}},
  \bibinfo{journal}{J. Mech. Phys. Solids}
  \textbf{\bibinfo{volume}{49}}, \bibinfo{pages}{761 } (\bibinfo{year}{2001}).

\bibitem[{\citenamefont{Kurganov et~al.}(2002)\citenamefont{Kurganov, Noelle,
  and Petrova}}]{KurganovSIAMJSC2002}
\bibinfo{author}{\bibfnamefont{A.}~\bibnamefont{Kurganov}},
  \bibinfo{author}{\bibfnamefont{S.}~\bibnamefont{Noelle}}, \bibnamefont{and}
  \bibinfo{author}{\bibfnamefont{G.}~\bibnamefont{Petrova}},
  \bibinfo{journal}{SIAM J. Sci. Comput.} \textbf{\bibinfo{volume}{23}},
  \bibinfo{pages}{707} (\bibinfo{year}{2002}).

\bibitem[{\citenamefont{Mignone et~al.}(2007)\citenamefont{Mignone, Bodo,
  Massaglia, Matsakos, Tesileanu, Zanni, and Ferrari}}]{MignoneApJS2007}
\bibinfo{author}{\bibfnamefont{A.}~\bibnamefont{Mignone}},
  \bibinfo{author}{\bibfnamefont{G.}~\bibnamefont{Bodo}},
  \bibinfo{author}{\bibfnamefont{S.}~\bibnamefont{Massaglia}},
  \bibinfo{author}{\bibfnamefont{T.}~\bibnamefont{Matsakos}},
  \bibinfo{author}{\bibfnamefont{O.}~\bibnamefont{Tesileanu}},
  \bibinfo{author}{\bibfnamefont{C.}~\bibnamefont{Zanni}}, \bibnamefont{and}
  \bibinfo{author}{\bibfnamefont{A.}~\bibnamefont{Ferrari}},
  \bibinfo{journal}{Astrophys. J. Suppl. S.}
  \textbf{\bibinfo{volume}{170}}, \bibinfo{pages}{228} (\bibinfo{year}{2007}).

\bibitem[{\citenamefont{Tan et~al.}(1994)\citenamefont{Tan, Zhang, Chang, and
  Zheng}}]{TanJDE1994}
\bibinfo{author}{\bibfnamefont{D.~C.} \bibnamefont{Tan}},
  \bibinfo{author}{\bibfnamefont{T.}~\bibnamefont{Zhang}},
  \bibinfo{author}{\bibfnamefont{T.}~\bibnamefont{Chang}}, \bibnamefont{and}
  \bibinfo{author}{\bibfnamefont{Y.~X.} \bibnamefont{Zheng}},
  \bibinfo{journal}{J. Differ. Equations}
  \textbf{\bibinfo{volume}{112}}, \bibinfo{pages}{1 } (\bibinfo{year}{1994}).

\bibitem[{\citenamefont{Shi et~al.}(2003)\citenamefont{Shi, Zhang, and
  Shu}}]{ShiJCP2003}
\bibinfo{author}{\bibfnamefont{J.}~\bibnamefont{Shi}},
  \bibinfo{author}{\bibfnamefont{Y.-T.} \bibnamefont{Zhang}}, \bibnamefont{and}
  \bibinfo{author}{\bibfnamefont{C.-W.} \bibnamefont{Shu}},
  \bibinfo{journal}{J. Comput. Phys.}
  \textbf{\bibinfo{volume}{186}}, \bibinfo{pages}{690 } (\bibinfo{year}{2003}).

\bibitem[{\citenamefont{Frisch}(1995)}]{Frisch1995}
\bibinfo{author}{\bibfnamefont{U.}~\bibnamefont{Frisch}},
  \emph{\bibinfo{title}{Turbulence : the legacy of A.N. Kolmogorov}}
  (\bibinfo{publisher}{Cambridge University Press}, \bibinfo{year}{1995}).

\bibitem[{\citenamefont{Limkumnerd and Sethna}(2008)}]{LimkumnerdJMPS2007}
\bibinfo{author}{\bibfnamefont{S.}~\bibnamefont{Limkumnerd}} \bibnamefont{and}
  \bibinfo{author}{\bibfnamefont{J.~P.} \bibnamefont{Sethna}},
  \bibinfo{journal}{J. Mech. Phys. Solids}
  \textbf{\bibinfo{volume}{56}}, \bibinfo{pages}{1450} (\bibinfo{year}{2008}).

\bibitem[{\citenamefont{Bardos and Lannes}(2010)}]{BardosArxiv2010}
\bibinfo{author}{\bibfnamefont{C.}~\bibnamefont{Bardos}} \bibnamefont{and}
  \bibinfo{author}{\bibfnamefont{D.}~\bibnamefont{Lannes}},
  \bibinfo{note}{e-print arXiv:1005.5329} (\bibinfo{year}{2010}).

\bibitem[{\citenamefont{Pumir and Siggia}(1992)}]{PumirPRL1992}
\bibinfo{author}{\bibfnamefont{A.}~\bibnamefont{Pumir}} \bibnamefont{and}
  \bibinfo{author}{\bibfnamefont{E.~D.} \bibnamefont{Siggia}},
  \bibinfo{journal}{Phys. Rev. Lett.} \textbf{\bibinfo{volume}{68}},
  \bibinfo{pages}{1511} (\bibinfo{year}{1992}).

\end{thebibliography}
\end{document}